\documentclass[12pt,sec]{article}
\usepackage{epsfig}
\usepackage{axodraw}
\begin{document}

\begin{center}
{\large  Problems of Double-Charm Production in $e^+e^-$ Annihilation at $\sqrt{s}=10.6$ GeV}\\[0.8cm]
{ Kui-Yong Liu$~^{(a)}$  Zhi-Guo He$~^{(a)}$ and~Kuang-Ta Chao$~^{(b,a)}$}\\[0.5cm]
{\footnotesize (a)~Department of Physics, Peking University,
 Beijing 100871, People's Republic of China}

{\footnotesize (b)~China Center of Advanced Science and Technology
(World Laboratory), Beijing 100080, People's Republic of China}
\end{center}

\begin{abstract}

Using the nonrelativistic QCD(NRQCD) factorization formalism, we
calculate the color-singlet cross sections for exclusive
production processes ${e^++e^-\rightarrow J/\psi+\eta_c}$~ and
~$e^++e^-\rightarrow J/\psi + \chi_{cJ}$~$(J=0,1,2)$ at the
center-of-mass energy $\sqrt{s}$=10.6 GeV.  The cross sections are
estimated to be $5.5$fb, $6.7$fb, $1.1$fb, and $1.6$fb for
$\eta_c, \chi_{c0}, \chi_{c1}$ and $\chi_{c2}$, respectively. The
calculated $J/\psi+\eta_c$ production rate is smaller than the
recent Belle data by about an order of magnitude, which might
indicate the failure of perturbative QCD calculation to explain
the double-charmonium production data. The complete
$\cal{O}$$(\alpha^2_{s})$ color-singlet cross section for
${e^++e^-\rightarrow \chi_{c0}+ c\bar {c} }$ is calculated. In
addition, we also evaluate the ratio of exclusive to inclusive
production cross sections. The ratio of $J/\psi\eta_c$ production
to $J/\psi c\bar{c}$ production could be consistent with the
experimental data.

PACS number(s): 12.40.Nn, 13.85.Ni, 14.40.Gx

\end{abstract}

Heavy quarkonium production is interesting in understanding both
perturbative and nonperturbative quantum chromodynamics (QCD). In
recent years the charmonium production has been studied in various
processes, such as in hadron-hadron collision, electron-proton
collision, fixed target experiments, $B$ meson decays, as well as
$Z^0$ decays. Among them, the study of charmonium production in
$e^+e^-$ annihilation is particularly interesting in testing the
quarkonium production mechanisms, the color-singlet model and the
color-octet model in the nonrelativistic QCD (NRQCD)\cite{bbl}
approach. This is not only because of the simpler parton structure
involved in this process, which may be helpful in reducing the
theoretical uncertainty, but also because of the spectacular
experimental prospect opened up by the two $B$ factories with
BaBar and Belle, which will allow a fine data analysis for
charmonium production with more than $10^8$ $e^+e^-$ annihilation
events in the continuum at $\sqrt{s}$=10.6 GeV.

Recently the Belle Collaboration has reported the observation of
prompt $J/\psi$ via double $c\bar{c}$ production from the
${e^+e^-}$ continuum\cite{belle}. For these results, not only the
large cross section ($\approx $0.9 pb) of the inclusive $J/\psi$
production due to the double $c\bar{c}$ is puzzling\cite{chao},
but also the exclusive production rate of $J/\psi\eta_c$,
$\sigma(e^+ + e^- \rightarrow J/\psi + \eta_c(\gamma))\times{\cal
B}(\eta_c\rightarrow \geq 4
charged)=(0.033_{-0.006}^{+0.007}\pm0.009)$pb, may not be
consistent with two previous calculations\cite{kane,brod}, which
gave a cross section of a few pb for $J/\psi \eta_c$. In fact,
recent perturbative QCD estimates of the $J/\psi c\bar{c}$ cross
section are only about 0.1 $\sim$ 0.2
pb\cite{cho,yuan,baek,kiselev}. So the calculations of exclusive
cross sections for $e^+e^-$ annihilation into $J/\psi \eta_c$ and
other double-charmonium states such as $J/\psi \chi_{cJ}(J=0,1,2)$
will be useful to clarify the problem. Experimentally, aside from
$e^+ + e^-\rightarrow J/\psi + \chi_{c0}$\cite{belle}, Belle
\cite{belle2} has also studied processes $e^+ + e^-\rightarrow
\chi_{c1}+X$ and $e^+ + e^-\rightarrow \chi_{c2}+X$, so we hope
that the double-charmonium production involving
$\chi_{cJ}(J=0,1,2)$ will be detectable in the near future. In the
following we will calculate the cross sections
$\sigma({e^++e^-\rightarrow J/\psi+\eta_c})$ and
$\sigma({e^++e^-\rightarrow J/\psi+\chi_{cJ}})$ in the leading
order perturbative QCD. To this order ($\sim\alpha_s^2$) the
color-singlet channel is dominant since all color-octet channels
are of high order of $v$, which is the relative velocity of the
charm quark and anti-charm quark in the charmonium, and therefore
suppressed relative to the color-singlet channel (the relative
suppression is at least of order $v^4$ for the cross sections).
Furthermore, in order to compare exclusive production with
inclusive production rates associated with the $\chi_{cJ}$
charmonium states, we will calculate the cross section
$\sigma({e^++e^-\rightarrow \chi_{c0}+c\bar{c}})$, and hope these
ratios will be useful for both inclusive and exclusive production
analyses at $\sqrt{s}$=10.6 GeV.

\begin{figure}[t]
\begin{center}
\vspace{-2.8cm}
\includegraphics[width=14cm,height=16cm]{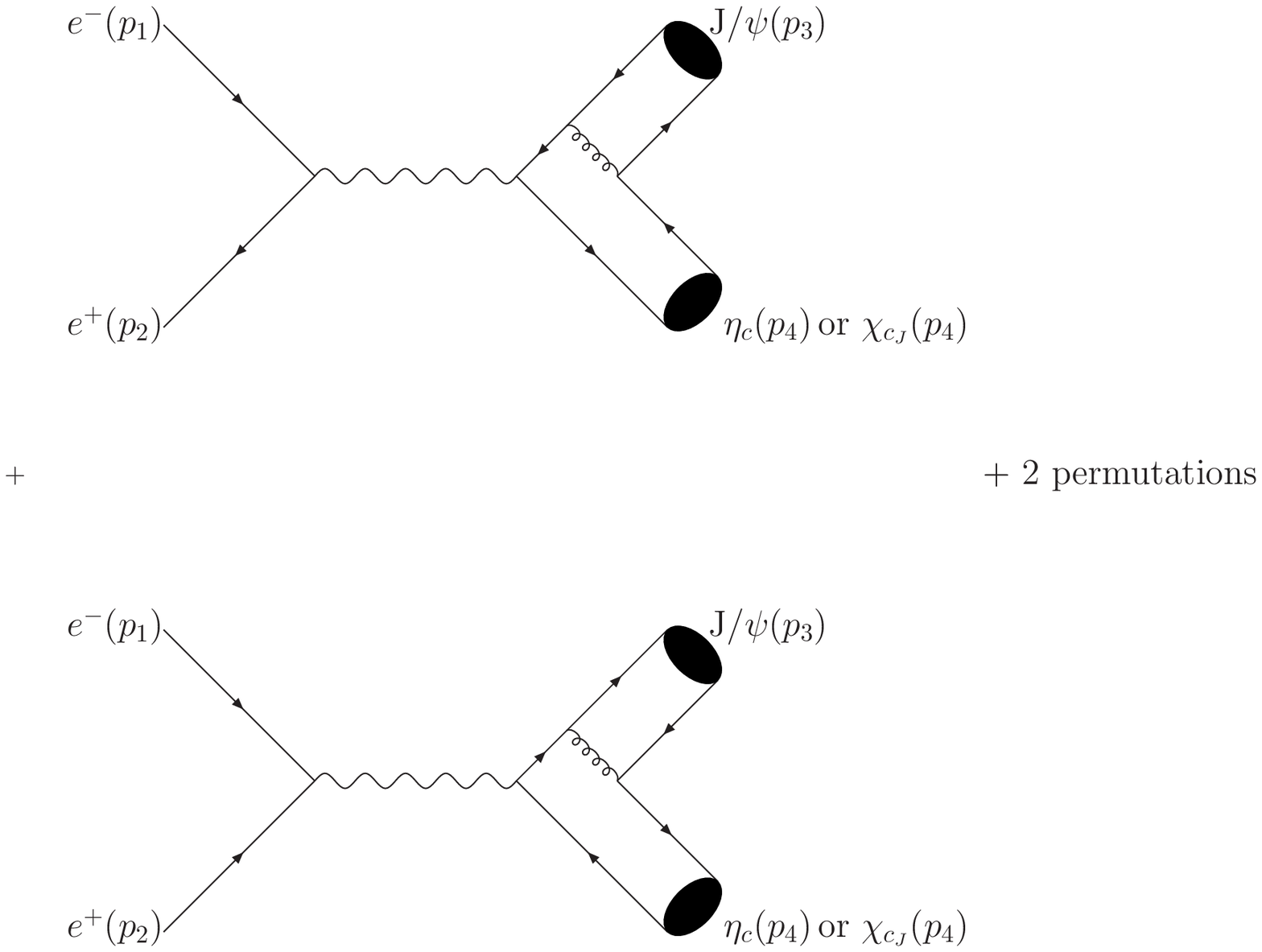}
\vspace{-4cm}
\end{center}
\caption{ Feynman diagrams for $e^+e^-\rightarrow J/\psi+\eta_c(
\chi_{cJ} )$.} \label{fig1}
\end{figure}
We now write down the scattering amplitude in the nonrelativistic
approximation to describe the creation of two color-singlet $c
\bar c$ pairs which subsequently hadronize to two charmonium
states in the $e^+e^-$ annihilation process in Fig.~\ref{fig1}
as\cite{cho2,pro}
\begin{eqnarray}
\label{amp2}   &&\hspace{-2.0cm}{\cal A}(a+b\rightarrow
Q\bar{Q}({}^{2S_\psi+1}L_{J_\psi})(p_3)+Q\bar{Q}({}^{2S
+1}L_{J})(p_4))= \sqrt{C_{L_\psi}}\sqrt{C_L}\sum\limits_{L_{\psi
z} S_{\psi z} }\sum\limits_{s_1s_2 }\sum\limits_{jk}
\sum\limits_{L_z S_z }\sum\limits_{s_3
s_4}\sum\limits_{il}\nonumber\\
&\times&\langle s_1;s_2\mid S_\psi S_{\psi z}\rangle \langle
L_\psi L_{\psi z };S_\psi S_{\psi z}\mid J_\psi J_{\psi
z}\rangle\langle 3j;\bar{3}k\mid
1\rangle\nonumber\\&\times&\langle s_3;s_4\mid S S_z\rangle\langle
L L_z ;S S_z\mid
J J_z\rangle\langle 3l;\bar{3}i\mid 1\rangle\nonumber\\
 &\times&\left\{
\begin{array}{ll}
{\cal A}(a+b\rightarrow
 Q_j(\frac{p_3}{2})+\bar{Q}_k(\frac{p_3}{2})+
 Q_l(\frac{p_4}{2})+\bar{Q}_i(\frac{p_4}{2}))&(L=S),\\
\epsilon^*_{\alpha}(L_Z) {\cal A}^\alpha(a+b\rightarrow
 Q_j(\frac{p_3}{2})+\bar{Q}_k(\frac{p_3}{2})+
 Q_l(\frac{p_4}{2})+\bar{Q}_i(\frac{p_4}{2}))
&(L=P),\\
\end{array}
\right.\nonumber\\
\end{eqnarray}
where $\langle 3j;\bar{3}k\mid 1\rangle =\delta_{jk}/\sqrt{N_c}$~,
$\langle 3l;\bar{3}i\mid 1\rangle=\delta_{li}/\sqrt{N_c}$~,~
$\langle s_1;s_2\mid S_\psi S_{\psi z}\rangle$~,~$\langle
s_3;s_4\mid S S_z\rangle$~,~ $\langle L_\psi L_{\psi z };S_\psi
S_{\psi z}\mid J_\psi J_{\psi z}\rangle$ and $ \langle L L_z ;S
S_z\mid J J_z\rangle$ are respectively the color-SU(3),
spin-SU(2), and angular momentum Clebsch-Gordan coefficients for
$Q\bar{Q}$ pairs projecting out appropriate bound states. ${\cal
A}(a+b\rightarrow Q_j(\frac{p_3}{2})+\bar{Q}_k(\frac{p_3}{2})+
 Q_l(\frac{p_4}{2})+\bar{Q}_i(\frac{p_4}{2}))$ is the scattering
 amplitude for double $Q\bar{Q}$ production and ${\cal A}^\alpha$
 is the derivative of the amplitude with respect to the relative
 momentum between the quark and anti-quark in the
bound state. The coefficients $C_{L_\psi}$ and $C_L$ can be
related to the radial wave function of the bound states or its
derivative with respect to the relative spacing as
\begin{equation}
\label{cs} C_s=\frac{1}{4\pi}|R_s (0)|^2,\ \ \
C_p=\frac{3}{4\pi}|R_p'(0)|^2.
\end{equation}
We introduce the spin projection operators $P_{SS_z}(p,q)$
as\cite{cho2,pro}
\begin{equation}
P_{SS_z}(p,q)\equiv\sum\limits_{s_1s_2 }\langle
s_1;s_2|SS_z\rangle
v(\frac{p}{2}-q;s_1)\bar{u}(\frac{p}{2}+q;s_2).
\end{equation}
Expanding $P_{SS_z}(P,q)$ in terms of the relative momentum $q$,
we get the projection operators and their derivatives, which will
be used in our calculation, as follows

\begin{equation}
\label{pjs} P_{1S_z}(p,0)=\frac{1}{2\sqrt{2}}\ \epsilon\!\!
/^*(S_z)(\not{p}+2m_c),
\end{equation}

\begin{equation}
\label{petc} P_{00}(p,0)=\frac{1}{2\sqrt{2}}\gamma_5(p\!\!\!\!
/+2m_c),
\end{equation}

\begin{equation}
\label{der} P_{1S_z}^{\alpha}(p,0)=\frac{1}{4\sqrt{2}m_c}
[\gamma^{\alpha}\not{\epsilon}^*(S_z)(\not{p}+2m_c)-
(\not{p}-2m_c)\not{\epsilon}(S_z)\gamma^{\alpha}].
\end{equation}

Then one can calculate the cross sections for the on-shell quarks
in the factorized form of NRQCD\cite{bbl}. The cross section for
${e^++e^-\rightarrow J/\psi+\eta_c}$ process in Fig.~\ref{fig1} is
given by
\begin{eqnarray}
\label{jsetc} \sigma(a(p_1)+b(p_2)\rightarrow
J/\psi(p_3)+\eta_c(p_4))=\frac{2\pi\alpha^2\alpha_s^2|R_s(0)|^4\sqrt{s-16m_c^2}
}{81m_c^2s^{3/2}}\int^1_{-1}|\bar{M}|^2 d\cos\theta,
\end{eqnarray}
where $\theta$ is the scattering angle between $\vec{p_1}$ and
$\vec{p_3}$, $|\bar{M}|^2$ is as follows
\begin{equation}
|\bar{M}|^2=\frac{16384m_c^2(t^2+u^2-32m_c^4)}{s^5}.
\end{equation}
The Mandelstam variables are defined as
\begin{equation}
s=(p_1+p_2)^2,
\end{equation}
\begin{equation}
t=(p_3-p_1)^2=4m_c^2-\frac{s}{2}(1-\sqrt{1-16m_c^2/s}\cos\theta),
\end{equation}
\begin{equation}
u=(p_3-p_2)^2=4m_c^2-\frac{s}{2}(1+\sqrt{1-16m_c^2/s}\cos\theta).
\end{equation}
The cross section for ${e^++e^-\rightarrow J/\psi+\chi_{cJ}}$
process is
\begin{eqnarray}
\label{jskc} &&\hspace{-3.0cm} \sigma(a(p_1)+b(p_2)\rightarrow
J/\psi(p_3)+\chi_{c_J}(p_4))=\nonumber
\\ && \frac{2\pi\alpha^2\alpha_s^2|R_s(0)|^2|R_p'(0)|^2\sqrt{s-16m_c^2}
}{27m_c^2s^{3/2}}\int^1_{-1}|\bar{M}_J|^2 d\cos\theta,
\end{eqnarray}
where  $|\bar{M}_J|^2$ for $\chi_{c0}$, $\chi_{c1}$ and
$\chi_{c2}$ are given by

\begin{eqnarray}
\label{kc0} &&\hspace{-1.1cm}
|\bar{M}_0|^2=2048(90112m_c^{10}-74752m_c^8t-74752m_c^8u+23360m_c^6t^2
+43136m_c^6tu+23360m_c^6u^2\nonumber\\&&-3152m_c^4t^3-7600m_c^4t^2u-
7600m_c^4tu^2-3152m_c^4u^3+162m_c^2t^4+444m_c^2t^3u\nonumber\\&&+564m_c^2t^2u^2
+444m_c^2tu^3+162m_c^2u^4-t^4u-3t^3u^2-3t^2u^3-tu^4)/(3s^7m_c^2),
\end{eqnarray}
\begin{eqnarray}
\label{kc1} &&\hspace{-2.5cm}
|\bar{M}_1|^2=32768(1792m_c^8+256m_c^6t+256m_c^6u-56m_c^4t^2
-64m_c^4tu-56m_c^4u^2-4m_c^2t^3\nonumber\\ &&
-20m_c^2t^2u-20m_c^2tu^2-4m_c^2u^3+t^4+2t^3u+2t^2u^2+2tu^3+u^4)/s^7,
\end{eqnarray}
\begin{eqnarray}
\label{kc2} &&\hspace{-1.5cm}
|\bar{M}_2|^2=4096(145408m_c^{10}-1024m_c^8t-1024m_c^8u-2368m_c^6t^2
-6400m_c^6tu-2368m_c^6u^2\nonumber\\
&&+16m_c^4t^3-208m_c^4t^2u-208m_c^4tu^2
+16m_c^4u^3+24m_c^2t^4+72m_c^2t^3u+96m_c^2t^2u^2\nonumber\\
&&+72m_c^2tu^3 +24m_c^2u^4-t^4u-3t^3u^2-3t^2u^3-tu^4)/(3s^7m_c^2).
\end{eqnarray}

 In the numerical calculations, we choose
$\sqrt{s}=10.6{\rm GeV},~~m_c=1.5{\rm GeV},~ \alpha_s=0.26$,
~$|R_s(0)|^2 =0.810{\rm GeV}^3$ and $|R_p'(0)|^2=0.075 {\rm
GeV^5}$\cite{wf}, and assume that in the nonrelativistic
approximation $m_{J/\psi}=m_{\eta_c}=m_{\chi_{cJ}}=2m_c$. The
numerical result for $e^+ + e^-\rightarrow J/\psi+\eta_c$ is
\begin{equation}
\label{jsetac} \sigma(e^+ + e^-\rightarrow J/\psi+\eta_c)=5.5
\rm{fb}.
\end{equation}

While the numerical result for the cross section of $e^+ +
e^-\rightarrow J/\psi+\eta_c$ is more than a factor of six smaller
than the experimental data\cite{belle} (with uncertainties due to
the unknown decay branching fractions into $\geq$ 4-charged
particles for the $\eta_c$), the calculated ratio of $\sigma(e^+ +
e^-\rightarrow J/\psi+\eta_c)/\sigma(e^+ + e^-\rightarrow J/\psi
+c\bar{c})\approx 0.037$ might be consistent with the experimental
result with the choice of $\sigma(e^+ + e^-\rightarrow J/\psi+
c\bar{c})=148$fb obtained by taking our input parameters.

The cross sections for $J/\psi\chi_{cJ}$ production at
$\sqrt{s}=10.6{\rm GeV}$ are given as
\begin{equation}
\label{jskc0} \sigma(e^+ + e^-\rightarrow J/\psi+\chi_{c0})=6.7
\rm{fb},
\end{equation}
\begin{equation}
\label{jskc1} \sigma(e^+ + e^-\rightarrow J/\psi+\chi_{c1})=1.1
\rm{fb},
\end{equation}
\begin{equation}
\label{jskc2} \sigma(e^+ + e^-\rightarrow J/\psi+\chi_{c2})=1.6
\rm{fb}.
\end{equation}

\begin{figure}[t]
\begin{center}
\includegraphics[width=12cm, height=10cm]{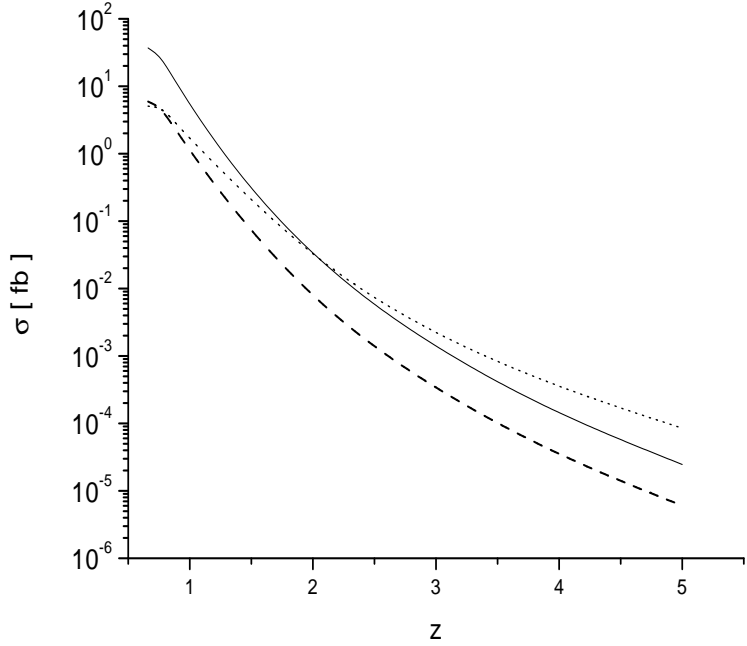}
\vspace{-1cm}
\end{center}
\caption{Cross sections for $\sigma(e^+ + e^-\rightarrow J/\psi+
\eta_c)$(solid line) and $\sigma(e^+ + e^-\rightarrow J/\psi+
\chi_{cJ})$(dashed line for $J=1$, dotted line for $J=2$) plotted
against the $e^+e^-$ center-of-mass energy $\sqrt{s}$ with
z=$\sqrt{s/s_0}$ and $\sqrt{s_0}=10.6 {\rm GeV}$.}
 \label{fig2}
\end{figure}

In Fig.~\ref{fig2}, we show the cross sections as functions of the
$e^+e^-$ center-of-mass energy $\sqrt{s}$, and we can see that the
cross sections for $e^+ + e^-\rightarrow J/\psi+\eta_c$, $e^+ +
e^-\rightarrow J/\psi+\chi_{c1}$ and $e^+ + e^-\rightarrow
J/\psi+\chi_{c2}$ decrease rapidly as $\sqrt{s}$ increases. But
the one with $J/\psi+\chi_{c2}$ decreases more slowly than that
with $J/\psi+\eta_c$ and $J/\psi+\chi_{c1}$.

At $\sqrt{s}=10.6 {\rm GeV}$ if we choose $\sigma(e^+ +
e^-\rightarrow \chi_{c1}+c\bar{c})=18.1$fb and $\sigma(e^+ +
e^-\rightarrow \chi_{c2}+c\bar{c})=8.4$fb which were obtained in
the fragmentation approximation in Ref.\cite{schuler}, then we
have the ratios
\begin{equation}
\label{ratio1} \frac{\sigma(e^+ + e^-\rightarrow
J/\psi+\chi_{c1})}{\sigma(e^+ + e^-\rightarrow
\chi_{c1}+c\bar{c})}=0.061,
\end{equation}
\begin{equation}
\frac{\sigma(e^+ + e^-\rightarrow J/\psi+\chi_{c2})}{\sigma(e^+ +
e^-\rightarrow \chi_{c2}+c\bar{c})}=0.19.
\end{equation}
\begin{equation}
\frac{\sigma(e^+ + e^-\rightarrow J/\psi+\chi_{c1})}{\sigma(e^+ +
e^-\rightarrow J/\psi+c\bar{c})}=0.007,
\end{equation}
\begin{equation}
\frac{\sigma(e^+ + e^-\rightarrow J/\psi+\chi_{c2})}{\sigma(e^+ +
e^-\rightarrow J/\psi+c\bar{c})}=0.011.
\end{equation}

As for the $\chi_{c0}$ inclusive double-charm production the rate
was not given in Ref.\cite{schuler}, we calculate $\sigma ( e^+ +
e^-\rightarrow \chi_{c0}+c\bar{c})$ in a complete form to the
$\cal O$$(\alpha_s^2)$ order in perturbative QCD.

\begin{figure}[t]
\begin{center}
\vspace{-2cm}
\includegraphics[width=12cm,height=16cm]{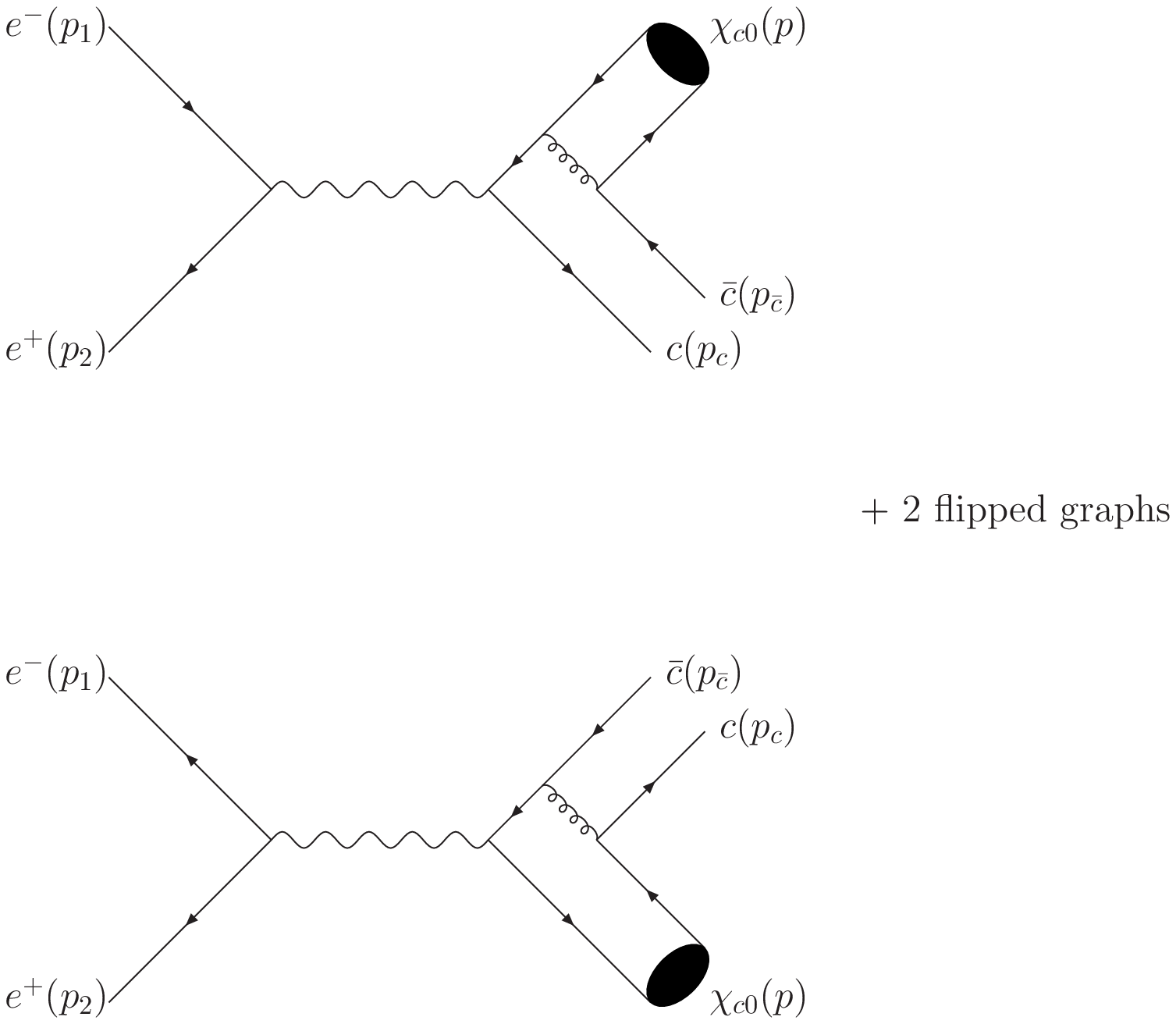}
\vspace{-5cm}
\end{center}
\caption{ Feynman diagrams for $e^++e^-\rightarrow
\chi_{c0}+c\bar{c}$ process.} \label{fig3}
\end{figure}

We give the amplitude of the first diagram in Fig.~\ref{fig3} for
$e^+ + e^-\rightarrow \chi_{c0}+c\bar{c}$ as
\begin{eqnarray}
&&\hspace{-2cm}
M=\sum\limits_{L_zS_z}\epsilon^*_{\sigma}(L_z)\langle
1L_z;1S_z|J=0,J_z=0\rangle\sqrt{C_L}\frac{ie_ceg_s^2[T^aT^a]_{li}}{\sqrt{3}}
\bar{v}(p_2)\gamma^{\mu}u(p_1)\frac{1}{s}\bar{u}_l(p_c)\nonumber
\\ & \times &[\gamma^{\alpha}P_{1S_z}\gamma_{\alpha}{\cal O}_{\mu}^{\sigma}+
\gamma^{\alpha}P^{\sigma}_{1S_z}\gamma_{\alpha}{\cal
O}_{\mu}]v_i(p_{\bar{c}}),
\end{eqnarray}
where $e_c=\frac{2}{3}e$ , $T^a$ is the $\rm{SU(3)}$ color matrix,
the matrix $\cal O$$_{\mu}$ is relevant to the on shell amplitude
and $\cal O$$_{\mu}^{\sigma}$ is its derivative with respect to
the relative momentum between the quarks that form the bound
state. We can also express the contributions of other three
diagrams in a similar way, and our numerical results are obtained
with the full contributions of these four diagrams. Some useful
information of the calculation is given in the Appendix.

We finally get the cross section for this process
\begin{equation}
\sigma(e^+ + e^-\rightarrow\chi_{c0}+c\bar{c})=49\rm{fb}.
\end{equation}
Then one has the ratio
\begin{equation}
\label{ratio} \frac{\sigma(e^+ + e^-\rightarrow
J/\psi+\chi_{c0})}{\sigma(e^+ + e^-\rightarrow
\chi_{c0}+c\bar{c})}=0.14,
\end{equation}
\begin{equation}
\label{ratio2} \frac{\sigma(e^+ + e^-\rightarrow
J/\psi+\chi_{c0})}{\sigma(e^+ + e^-\rightarrow
J/\psi+c\bar{c})}=0.045.
\end{equation}

\begin{figure}[t]
\begin{center}
\vspace{-1.0cm}
\includegraphics[width=12cm,height=10cm]{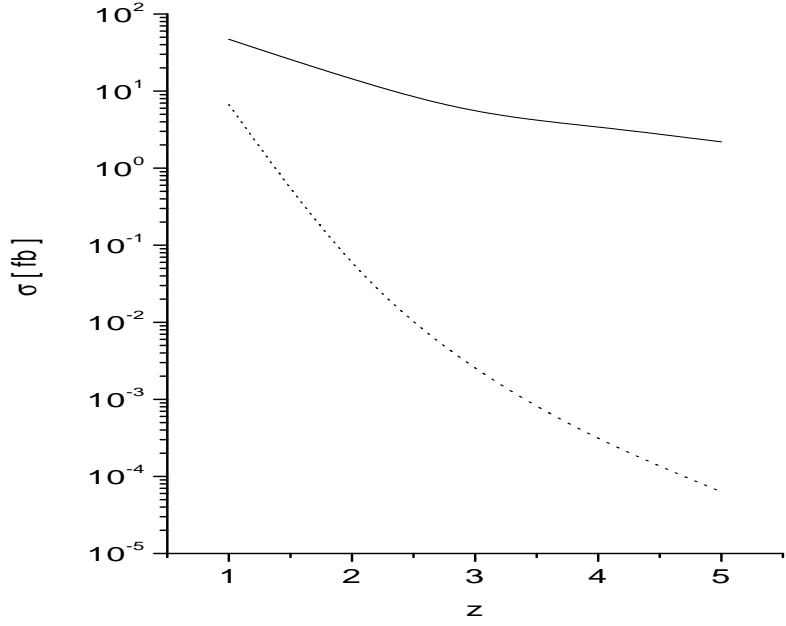}
\vspace{-1cm}
\end{center}
\caption{cross sections for $\sigma(e^+ + e^-\rightarrow
\chi_{c0}+ c\bar{c})$(solid line) and $\sigma(e^+ + e^-\rightarrow
J/\psi+\chi_{c0})$(dotted line) plotted against the $e^+e^-$
center-of-mass energy $\sqrt{s}$ with z=$\sqrt{s/s_0}$ and
$\sqrt{s_0}=10.6 {\rm GeV}$.} \label{fig4}
\end{figure}

In Fig.~\ref{fig4}, we show cross sections for $\sigma(e^+ +
e^-\rightarrow \chi_{c0}+c\bar{c})$(solid line) and $\sigma(e^+ +
e^-\rightarrow J/\psi+\chi_{c0})$(dotted line) plotted against the
$e^+e^-$ center-of-mass energy $\sqrt{s}$ with z=$\sqrt{s/s_0}$
and $\sqrt{s_0}=10.6 {\rm GeV}$. One can see the ratio in
Eq.~(\ref{ratio}) decreases drastically as the center-of-mass
energy increases. This is consistent with the result in
Fig.~\ref{fig2}. We hope the ratios between Eq.~(\ref{ratio1}) and
Eq.~(\ref{ratio2}) could be tested in the near future.

In summary, despite of many uncertainties due to the relativistic
corrections, the QCD radiative corrections, the possible
color-octet channel contributions, and the choice of physical
parameters (e.g. the charm quark mass and the strong coupling
constant), both the inclusive and exclusive double charm
production cross sections calculated in perturbative QCD turned
out to be seriously underestimated as compared with data.
Therefore we intend to conclude, as in \cite{chao}, that it seems
hard to explain the double charm production data observed by Belle
based on perturbative QCD (including both color-singlet and
color-octet channels), and possible nonperturbative QCD effects
should be considered at $\sqrt{s}=10.6 {\rm GeV}$.

While we were about to submit our result, there appeared one paper
which also considered exclusive double-charmonium
production\cite{bl}. Those authors took the QED effects into
account in addition to the QCD effects that we considered.  We
find our result for the exclusive double-charmonium production is
consistent with theirs but we also analyzed some inclusive
processes which were not discussed in Ref.~\cite{bl}.

\section*{Acknowledgments}
The authors thank L.K. Hao and Z.Z. Song for useful discussions.
We also thank C.F. Qiao for helpful comments. This work was
supported in part by the National Natural Science Foundation of
China, and the Education Ministry of China.

\section*{Appendix}

In this appendix we give the cross section for the $e^+ +
e^-\rightarrow \chi_{c0}+c\bar{c}$ process shown in
Fig.~\ref{fig3}.
\begin{equation}
d\sigma=\frac{|\bar{M}|^2}{2s(2\pi)^5}\delta^4(p_1+p_2-p_c-p_{\bar{c}}-p)
\frac{d^3p_c}{2E_c}\frac{d^3p_{\bar{c}}}{2E_{\bar{c}}}\frac{d^3p}{2E}.
\end{equation}
It is convenient to rewrite the cross section as
\begin{eqnarray}
&&\hspace{-0.6cm}  d\sigma=
\frac{|\bar{M}|^2}{2s(2\pi)^5}\delta^4(p_1+p_2-\eta-p)\delta^4(\eta-p_c-p_{\bar{c}})
\frac{d^3p_c}{2E_c}\frac{d^3p_{\bar{c}}}{2E_{\bar{c}}}\frac{d^3p}{2E}d^4\eta\nonumber
\\ &&=\frac{|\bar{M}|^2}{2s(2\pi)^5}\delta^4(p_1+p_2-\eta-p)\delta^4(\eta-p_c-p_{\bar{c}})
\frac{d^3p_c}{2E_c}\frac{d^3p_{\bar{c}}}{2E_{\bar{c}}}\frac{d^3p}{2E}\frac{d^3\eta}{2E_\eta}dm_{\eta}^2,\nonumber\\
\end{eqnarray}
where $m_{\eta}^2=\eta^2$.

The integral over the phase-space of $c\bar{c}$ is evaluated in
the corresponding center-of-mass frame
\begin{eqnarray}
\frac{d^3p_c'}{2E_c'}\frac{d^3p_{\bar{c}}'}{2E_{\bar{c}}'}\delta^4(\eta'-p_c'-p_{\bar{c}}')=
\frac{1}{8m_{\eta}^2}\lambda^{1/2}(m_{\eta}^2,m_c^2,m_c^2)d\Omega',
\end{eqnarray}
where $\lambda(a^2,b^2,c^2)=a^4+b^4+c^4-2a^2b^2 -2a^2c^2-2b^2c^2$.

The remaining integration are performed in the $e^+e^-$
center-of-mass frame
\begin{eqnarray}
\frac{d^3p}{2E}\frac{d^3\eta}{2E_{\eta}}\delta^4(p_1+p_2-\eta-p)=
\frac{1}{8s}\lambda^{1/2}(s,m_{\eta}^2,m_p^2)d\Omega,
\end{eqnarray}
where $m_p=2m_c$, is the mass of the bound state.

Finally we have
\begin{equation}
d\sigma=\frac{|\bar{M}|^2C_P}{64s^2(2\pi)^5m_{\eta}m_c}\lambda^{1/2}(m_{\eta}^2,m_c^2,m_c^2)
\lambda^{1/2}(s,m_{\eta}^2,m_p^2)d\Omega'd\Omega dm_{\eta}.
\end{equation}

The limit of $m_{\eta}$ is
\begin{equation}
2m_c\leq m_{\eta}\leq \sqrt{s}-m_p.
\end{equation}

To accomplish the integration we use the Lorentz transformation
between the two frames as $L=R_2R_1$, where

\begin{eqnarray}
R_1=\left ( \begin{array}{cccc}
  \sqrt{1+\frac{\vec{p}^2}{m_{\eta}^2}} & 0 & 0 &-\frac{|\vec{p}|}{m_{\eta}}  \\
  0 & 1 & 0 & 0 \\
  0 & 0 & 1 & 0 \\
  -\frac{|\vec{p}|}{m_{\eta}} & 0 & 0 & \sqrt{1+\frac{\vec{p}^2}{m_{\eta}^2}}
\end{array} \right ),
\end{eqnarray}

\begin{eqnarray}
R_2=\left ( \begin{array}{cccc}
  1 & 0 & 0 & 0 \\
  0 & \cos\theta & 0 & \sin\theta \\
  0 & 0 & 1 & 0 \\
  0 & -\sin\theta & 0 & \cos\theta
\end{array} \right ).
\end{eqnarray}

The momenta in the $e^+e^-$ center-of-mass frame are

\begin{equation}
p_1=(\sqrt{s}/2, 0, 0, \sqrt{s}/2),
\end{equation}

\begin{equation}
p_2=(\sqrt{s}/2, 0, 0, -\sqrt{s}/2),
\end{equation}

\begin{equation}
p_c=R_2R_1p_c',
\end{equation}

\begin{equation}
p_{\bar{c}}=R_2R_1p_{\bar{c}}',
\end{equation}

\begin{equation}
p=(\sqrt{\vec{p}^2+m_p^2}, |\vec{p}|\sin\theta,0,
|\vec{p}|\cos\theta),
\end{equation}
where $p_c'$ and $p_{\bar{c}}'$ are the momenta of $c$ and $
\bar{c}$ in the $\Omega'$ frame, and they are

\begin{equation}
p_c'=(E_c', |\vec{p_c}'|\sin\theta'\cos\theta',
|\vec{p_c}'|\sin\theta'\cos\theta', |\vec{p_c}'|\cos\theta'),
\end{equation}

\begin{equation}
p_{\bar{c}}'=(E_{\bar{c}}', -|\vec{p_c}'|\sin\theta'\cos\theta',
-|\vec{p_c}'|\sin\theta'\cos\theta', -|\vec{p_c}'|\cos\theta').
\end{equation}
In Fig.~\ref{fig3} the lower (non-fragmentation) diagrams give
very small contributions (about 3 percent), so for simplicity here
we only write down the expressions for the contribution of the
upper diagrams and give

\begin{equation}
|\bar{M}|^2=\frac{2(4\pi)^4\alpha^2\alpha_s^2}{27}(aa+2ab+bb).
\end{equation}
We define $pp_1=p.p_1$, $pp_2=p.p_2$, $pp_3=p.p_c$,
$pp_4=p.p_{\bar{c}}$, $p_{13}=p_1.p_c$, $p_{14}=p_1.p_{\bar{c}}$,
$p_{23}=p_2.p_c$, $p_{24}=p_2.p_{\bar{c}}$. We notify $aa=bb$ and
\begin{eqnarray}
&& \hspace{-0.5cm}aa = [4(800m_c^{10}s + 800m_c^8p_{14}pp_2 +
800m_c^8p_{24}pp_1+1440m_c^8pp_3s +160m_c^6p_{13}p_{24}pp_3
\nonumber
\\ &&+ 160m_c^6p_{14}p_{23}pp_3 + 1440m_c^6p_{14}pp_2pp_3 + 1440m_c^6p_{24}pp_1pp_3 + 900m_c^6pp_3^2s
\nonumber
\\ &&+ 184m_c^4p_{13}p_{24}pp_3^2 + 184m_c^4p_{14}p_{23}pp_3^2 + 856m_c^4p_{14}pp_2pp_3^2 +
856m_c^4p_{24}pp_1pp_3^2
 \nonumber
\\ &&+ 216m_c^4pp_3^3s + 56m_c^2p_{13}p_{24}pp_3^3 + 56m_c^2p_{14}p_{23}pp_3^3 + 168m_c^2p_{14}pp_2pp_3^3
\nonumber
\\ &&+ 168m_c^2p_{24}pp_1pp_3^3 + 13m_c^2pp_3^4s + 2p_{13}p_{24}pp_3^4+ 2p_{14}p_{23}pp_3^4)]/[3m_c^2s^2(64m_c^{12}
\nonumber
\\ && +
192m_c^{10}pp_3+ 240m_c^8pp_3^2 + 160m_c^6pp_3^3 + 60m_c^4pp_3^4 +
12m_c^2pp_3^5 + pp_3^6)],
\end{eqnarray}

\begin{eqnarray}
&&\hspace{-0cm} ab= [4(400m_c^{10}s + 400m_c^8p_{13}pp_2 +
400m_c^8p_{14}pp_2 + 400m_c^8p_{23}pp_1+ 400m_c^8p_{24}pp_1
\nonumber
\\ &&+ 400m_c^8pp_1pp_2 + 480m_c^8pp_3s +
400m_c^8pp_{34}s + 480m_c^8pp_4s + 80m_c^6p_{13}p_{24}pp_3
\nonumber
\\ && + 80m_c^6p_{13}p_{24}pp_4 + 280m_c^6p_{13}pp_2pp_3 + 440m_c^6p_{13}pp_2pp_4 + 80m_c^6p_{14}p_{23}pp_3
 \nonumber
\\ &&+ 80m_c^6p_{14}p_{23}pp_4 + 440m_c^6p_{14}pp_2pp_3 + 280m_c^6p_{14}pp_2pp_4 + 280m_c^6p_{23}pp_1pp_3
\nonumber
\\ &&+ 440m_c^6p_{23}pp_1pp_4 + 440m_c^6p_{24}pp_1pp_3+ 280m_c^6p_{24}pp_1pp_4 + 240m_c^6pp_1pp_2pp_3
 \nonumber
\\ &&- 400m_c^6pp_1pp_2pp_{34} + 240m_c^6pp_1pp_2pp_4 + 140m_c^6pp_3^2s + 240m_c^6pp_3pp_{34}s
 \nonumber
\\ &&+ 476m_c^6pp_3pp_4s + 240m_c^6pp_{34}pp_4s + 140m_c^6pp_4^2s + 40m_c^4p_{13}p_{24}pp_3^2
+ 104m_c^4p_{13}p_{24}pp_3pp_4 \nonumber
\\ && + 40m_c^4p_{13}p_{24}pp_4^2+
20m_c^4p_{13}pp_2pp_3^2 + 288m_c^4p_{13}pp_2pp_3pp_4 +
120m_c^4p_{13}pp_2pp_4^2 \nonumber
\\ &&+ 40m_c^4p_{14}p_{23}pp_3^2 + 104m_c^4p_{14}p_{23}pp_3pp_4 + 40m_c^4p_{14}p_{23}pp_4^2
+ 120m_c^4p_{14}pp_2pp_3^2\nonumber
\\ && + 288m_c^4p_{14}pp_2pp_3pp_4 +
20m_c^4p_{14}pp_2pp_4^2+ 20m_c^4p_{23}pp_1pp_3^2
 + 288m_c^4p_{23}pp_1pp_3pp_4  \nonumber
\\ &&+ 120m_c^4p_{23}pp_1pp_4^2+ 120m_c^4p_{24}pp_1pp_3^2 + 288m_c^4p_{24}pp_1pp_3pp_4
 + 20m_c^4p_{24}pp_1pp_4^2\nonumber
\\ &&  - 240m_c^4pp_1pp_2pp_3pp_{34} + 144m_c^4pp_1pp_2pp_3pp_4 - 240m_c^4pp_1pp_2pp_{34}pp_4
  + 108m_c^4pp_3^2pp_4s\nonumber
\\ && + 144m_c^4pp_3pp_{34}pp_4s + 108m_c^4pp_3pp_4^2s + 28m_c^2p_{13}p_{24}pp_3^2pp_4
+ 28m_c^2p_{13}p_{24}pp_3pp_4^2\nonumber
\\ &&+ 12m_c^2p_{13}pp_2pp_3^2pp_4 +
72m_c^2p_{13}pp_2pp_3pp_4^2 + 28m_c^2p_{14}p_{23}pp_3^2pp_4
 + 28m_c^2p_{14}p_{23}pp_3pp_4^2 \nonumber
\\ &&+ 72m_c^2p_{14}pp_2pp_3^2pp_4 + 12m_c^2p_{14}pp_2pp_3pp_4^2
+ 12m_c^2p_{23}pp_1pp_3^2pp_4 + 72m_c^2p_{23}pp_1pp_3pp_4^2
\nonumber
\\ &&+ 72m_c^2p_{24}pp_1pp_3^2pp_4+ 12m_c^2p_{24}pp_1pp_3pp_4^2
- 144m_c^2pp_1pp_2pp_3pp_{34}pp_4 + 13m_c^2pp_3^2pp_4^2s \nonumber
\\ &&+ 2p_{13}p_{24}pp_3^2pp_4^2 + 2p_{14}p_{23}pp_3^2pp_4^2)]/[3m_c^2s^2(64m_c^{12}
+ 96m_c^{10}pp_3 + 96m_c^{10}pp_4+ 48m_c^8pp_3^2 \nonumber
\\ && + 144m_c^8pp_3pp_4 +
48m_c^8pp_4^2 + 8m_c^6pp_3^3 + 72m_c^6pp_3^2pp_4 +
72m_c^6pp_3pp_4^2 +8m_c^6pp_4^3\nonumber
\\ &&+ 12m_c^4pp_3^3pp_4+ 36m_c^4pp_3^2pp_4^2 + 12m_c^4pp_3pp_4^3
+ 6m_c^2pp_3^3pp_4^2 + 6m_c^2pp_3^2pp_4^3 + pp_3^3pp_4^3)].
\end{eqnarray}

\end{document}